\documentclass[twocolumn,prl,showpacs,superscriptaddress]{revtex4-1}
\usepackage{graphicx}
\usepackage{amsmath}
\usepackage{amssymb}

\begin{document}
\title{Promoting cold-start items in recommender systems}

\author{Jin-Hu Liu}
\affiliation{Web Sciences Center, University of Electronic Science and Technology of China, Chengdu 610054, People's Republic of China}
\author{Tao Zhou}
\email{E-mail: zhutou@ustc.edu}
\affiliation{Web Sciences Center, University of Electronic Science and Technology of China, Chengdu 610054, People's Republic of China}
\author{Zi-Ke Zhang}
\affiliation{Alibaba Research Center for Complexity Sciences, Hangzhou Normal University, Hangzhou, 311121, People's Republic of China}
\author{Zimo Yang}
\affiliation{Web Sciences Center, University of Electronic Science and Technology of China, Chengdu 610054, People's Republic of China}
\affiliation{Alibaba Group, Hangzhou, 311100, People's Republic of China}
\author{Chuang Liu}
\affiliation{Alibaba Research Center for Complexity Sciences, Hangzhou Normal University, Hangzhou, 311121, People's Republic of China}
\author{Wei-Min Li}
\affiliation{Web Sciences Center, University of Electronic Science and Technology of China, Chengdu 610054, People's Republic of China}
\affiliation{Beijing Baifendian Information Technology Co., Ltd., Beijing, 100080, People's Republic of China}

\date{\today}

\begin{abstract}
As one of major challenges, cold-start problem plagues nearly all recommender systems. In particular, new items will be overlooked, impeding the development of new products online. Given limited resources, how to utilize the knowledge of recommender systems and design efficient marketing strategy for new items is extremely important. In this paper, we convert this ticklish issue into a clear mathematical problem based on a bipartite network representation. Under the most widely used algorithm in real e-commerce recommender systems, so-called the item-based collaborative filtering, we show that to simply push new items to active users is not a good strategy. To our surprise, experiments on real recommender systems indicate that to connect new items with some less active users will statistically yield better performance, namely these new items will have more chance to appear in other users' recommendation lists. Further analysis suggests that the disassortative nature of recommender systems contributes to such observation. In a word, getting in-depth understanding on recommender systems could pave the way for the owners to popularize their cold-start products with low costs.

\end{abstract}


\maketitle

\section{Introduction}
Thanks to the blazing development of Internet, e-commerce has flourished over the past decades. With the online buy-and-sell platforms getting increasingly more available products (e.g., more than a billion products in \emph{taobao.com}), shopping online has become a fashionable style of living and more people choose to purchase on the Internet rather than go to a store. E-commerce makes our life much more convenient, meanwhile, it throws us into a dilemma of information overloads. Facing millions of items online, finding out favourites is rather difficult. As an effective information filtering tool, recommender system is thus of particular significance nowadays \cite{Ricci2011,Lv2012}. In fact, it has already made considerable contributions to the socioeconomic fields in the past decade. For example, 60\% of DVDs rented by \emph{Netflix} are selected based on personalized recommendations, and about a half of sales in \emph{Amazon} are brought by recommendations \cite{Lv2012}. Consequently, recommender systems have received huge attentions from both physicists and computer scientists, and many advanced recommendation algorithms are proposed recently, including collaborative filtering \cite{Goldberg1992, Shardanand1995, Sarwar2001, Deshpande2004, Schafer2007, KimheungNam2010}, content-based analysis \cite{Balabanovic1997, Pazzani2007, Katja2013}, dimensionality reduction techniques \cite{Hofmann2004, Takacs2007, Bunte2012}, diffusion-based methods \cite{Huangzan2004,Zhangyc2007prl,Zhangyc2007epl,PRE07zhoutao,PNASzhoutao,Liujg2011,LLY2011,Zhouyanbo2013}, and so on.

One long-standing challenge, called cold-start problem, has plagued almost all recommender systems. Namely, when new users or items enter the system, there is usually insufficient information to produce reasonable recommendation \cite{Schein2002}. Considering this fact, several potential solutions have been raised. The additional content information \cite{Schein2002,Par2009,Cantador2010,Ronen2013}, tagging information \cite{Zhangzk20102,Zhangzk20122,Yin2013} and cross-domain information \cite{Zhangl2013} can be used to marginally relieve this problem, but they don't work in a purely cold-start setting, where no information is available to form any basis for recommendations. Furthermore, improving diversity and novelty of recommended lists can help new items be pushed out \cite{Ziegler2005,PNASzhoutao,Zhoutao09NJP}.

Practically speaking, as a holder of the recommender system, one can ask for extra information to generate initial profiles on users or items \cite{Par2009}, or probe users' preferences by pushing to them some carefully selected items according to complicated algorithms \cite{Zhoujunlin2011}. Both methods are costly and risky. In contrast, an owner would like to popularize her new items. An improper method, called ``shilling attacks", injects a number of mendacious users into the system to raise predicted ratings of new items, and thus enhances the possibility of these new item to appear in the recommendation lists \cite{Lam2004,Mobasher2005}. But, it is easily to be detected \cite{Mobasher2006,Mehta2007,Mobasher2007}. Furthermore, as a wide-spreading market strategy, advertisements are generally preferred and become more and more prosperous \cite{Verhoef2009}. However, to popularize new items costs a lot and imposes an unbearable financial burden for small businesses \cite{Edward2008}. As mentioned above, how to promote new items under limited marketing resources is a nontrivial challenge and the knowledge of recommendation algorithm may be helpful. Putting aside operational details, if the marketing activities can bring some purchases of certain users, a smart marketing manager will carefully chose the target users so that these purchases can lead to more exposures in the recommendation lists afterwards.

Taking a stand as a marketing manager, in this paper, we focus on how to promote cold-start items by utilizing the knowledge of recommender systems. The main contributions are threefold: (i) We convert this ticklish problem into a clear mathematical model that ignores some insignificant details. (ii) We show that to push new items to active users, a straightforward strategy that will jump into our mind at the first time, is to our surprise a poor-performed strategy. (iii) We propose a degree-based solution that outperforms some baseline methods.

\section{Results}
\begin{figure}[!ht]
\begin{center}
\includegraphics[width=8cm]{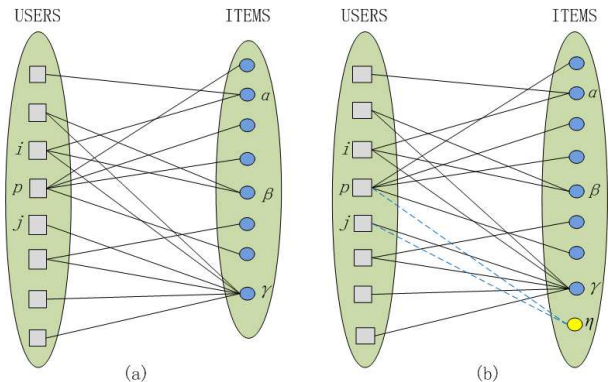}
\caption{\textbf{How to add a cold-start item to the user-item bipartite network.} Users and items are represented by squares and circles respectively, and solid lines represent the existent links between them. Plot (a) is the original network, and plot (b) is the network after adding the item $\eta$ (the yellow circle). The dotted lines are new links connecting $\eta$ with two existent users.}\label{Fig1}
\end{center}
\end{figure}

Recommendation can be considered as a variant of link prediction in bipartite networks \cite{lvlinyuan2011} and thus the better understanding of network structures can in principle improve the quality of recommendations \cite{Huang2007, Koren2011, Zhangqm2013,Weizeng2014}. We denote a recommender system by a user-item bipartite network $G(U,O,E)$, where $U=\{U_1,U_2,\ldots,U_n\}$ and $O=\{O_1,O_2,\ldots,O_m\}$ are respectively the sets of users and items, and $E$ is the set of links connecting users and items. Consequently, we use the adjacent matrix, $A$, to describe the user-item relations: if user $U_i$ has purchased item $O_\alpha$, $a_{i\alpha}=1$, otherwise $a_{i\alpha}=0$ (throughout this paper we use Latin and Greek letters, respectively, for user- and item-related indices). Figure \ref{Fig1}(a) illustrates a small bipartite network that consists of eight users (gray squares) and eight items (blue circles). $k_i$, the degree of user $U_i$, is defined as the number of items linked to $U_i$. Analogously, the degree of item $O_\alpha$, denoted by $k_\alpha$ is the number of users connected to $O_\alpha$. For example, as shown in Figure \ref{Fig1}(a), $k_i=3, k_j=1$ and $k_\alpha=2$. The user degree distribution $P_u(k)$, is the probability that the degree of a randomly selected user, is equal to $k$. The item degree distribution $P_o(k)$ is defined in a similar way. Degree distribution reflects the network heterogeneity \cite{Barabasi199901}.

We consider two real data sets with anonymous users in this paper, including (a) \emph{Tmall.com} (TM): an open business-to-consumer (B2C) platform where enrolled businessmen can sell legal items to customers; (b) \emph{Coo8.com} (Coo8): a well established online retailer mainly trading in electrical household appliances and a leading supplier to daily necessities. In order to avoid the isolate nodes in the data sets, each user has bought at least one item, and each item has been purchased at least once. Table \ref{tab:data} shows the basic statistics of the two data sets. Due to the different types of products, these networks have much different average item degrees. As shown in Figure 2, all degree distributions are heavy-tailed and the item degree distributions are generally more heterogenous than the corresponding user degree distributions. These observations complement previous empirical analyses on user-item bipartite networks \cite{Lambiotte2005,Goncalves2008,Zimo2012,Shangmz2010}.

\begin{table}[h]
\centering \caption{Basic statistical properties of the two data sets. $n$, $m$, and $w$ represent the number of users, items and links, $\langle k_{user} \rangle$ and $\langle k_{item} \rangle$ stand for the average degrees of users and items, and $S=\frac{w}{n\times m}$ denotes the data sparsity.}
\setlength\tabcolsep{2pt}
\begin{tabular}{ccccccc}
\hline
\hline
Data & $n$ & $m$ & $w$ & $\langle k_{user} \rangle$ & $\langle k_{item} \rangle$ & $S$ \\
\hline
\emph{TM} & 103,867 & 83,342 & 113,624 & 1.09 & 1.36 & 1.31 $\times10^{-5}$\\
\hline
\emph{Coo8} & 77,947 & 18,751 & 94,457 & 1.21 & 5.04 & 6.46 $\times10^{-5}$\\
\hline
\hline
\end{tabular}\label{tab:data}
\end{table}

The nearest neighbors' degree for user $U_i$, denoted by $d_{nn}^u(i)$, is defined as the average degree over all items connected to $U_i$ \cite{Shangmz2010}. For example, in Figure \ref{Fig1}(a), $d_{nn}^u(i)=\frac{k_\alpha+k_\beta+k_\gamma}{k_i}=\frac{10}{3}$. Furthermore, the degree-dependent nearest neighbors' degree, $\langle d_{nn}^u(k)\rangle$ is the average nearest neighbors' degree over all users of degree $k$, namely $\langle d_{nn}^u(k) \rangle= \langle d_{nn}^u(i) \rangle _{k_i=k}$. Corresponding concepts for items, $P_o(k)$, $d_{nn}^o(k)$ and $\langle d_{nn}^o(k) \rangle $, are defined in a similar way and thus omitted here \cite{Shangmz2010}. The degree-dependent nearest neighbors' degree is an appropriate index to characterize the network assortativity \cite{Poster2001}. As shown in Figure 2, both the two networks are disassortative.

\begin{figure}[!ht]
\begin{center}
\includegraphics[width=10cm]{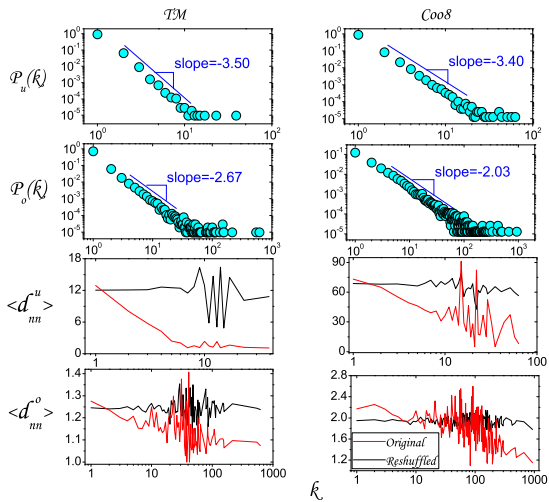}
\caption{\textbf{Degree distributions and degree correlations.} All degree distributions are power-law-like, with exponents being estimated by the maximum likelihood methods \cite{Goldstein2004,Clauset2009}. $\langle d_{nn}^u \rangle$ and $\langle d_{nn}^o \rangle$ are respectively showed in the 3rd and 4th rows, where red and black lines representing the results from original and reshuffled networks. Results of reshuffled networks are obtained by averaging over five independent realizations.}\label{Fig2}
\end{center}
\end{figure}

Recommender systems typically produce a given-length list of unpurchased items for each user based on his historical purchases. Of nothing comes nothing, that is to say, it is impossible to predict links for an isolate user or item. So only after having been purchased by some users, an item could have the chance to appear in some other users' recommendation lists. In real e-commerce web sites, to get a new customer is highly costly, and thus under the limited investment, choosing users with considerable coming influence on further recommendations is absolutely critical. Concretely speaking, this problem is described as follow. Given a bipartite network containing $n$ users, $m$ items and $w$ links. A novel item $O_\eta$ enters this network, and it can at most establish $R$ links to users. Given the recommendation algorithm, we need to answer the question that how to choose such $R$ users to maximize the frequency that $O_\eta$ appears in other $(n-R)$ users' recommendation lists. For example, in Figure \ref{Fig1}(b), item $O_\eta$ (the yellow circle) comes and needs to link to some existent users. If $R=1$, then to choose which user, $U_p$ (most active user), $U_j$ (one of the most inactive users) or another one, can make $O_\eta$ be recommended more times?

We consider four strategies to choose those $R$ users: (I) Maximum-degree strategy (MaxD). To rank all users in the descending order of degree, and select the top-$R$ users, where users with the same degree are ranked randomly. (II) Minimum-degree strategy (MinD). To rank all users in the ascending order of degree, and select top-$R$ users, where users with the same degree are ranked randomly. (III) Preferential attachment strategy (PA). Each user's probability to be selected is proportional to her degree. (IV) Random strategy (RAN). The $R$ users are selected completely randomly. Actually, all strategies above can be unified by a selecting probability of any user $U_i$, $p(U_i)\propto\frac{k_i^\tau}{\sum{k_i^\tau}}$, where $\tau$ is a tunable parameter. More specifically, the strategies MaxD, MinD, PA and RAN correspond to the cases of $\tau=+\infty$, $\tau=-\infty$, $\tau=1$ and $\tau=0$, respectively.

Among existent recommendation algorithms, item-based collaborative filtering (ICF) has found the widest applications in real e-commerce platforms for its accuracy, stability, scalability and robustness \cite{Sarwar2001,Deshpande2004,Mobasher2007}. Here, we apply cosine similarity for each pair of items, say
\begin{equation}
sim(\alpha,\beta)=\sum_{s=1}^n \frac{a_{s \alpha} a_{s \beta}}{\sqrt{k_\alpha k_\beta}},\label{eq:e3}
\end{equation}
where $k_\alpha$ and $k_\beta$ are degrees of items $O_\alpha$ and $O_\beta$, respectively. For the target user $U_i$, we calculate the accumulative score $w_{i\alpha}$ for each item $O_\alpha$ by
\begin{equation}
w_{i\alpha}=\sum_{\gamma \neq \alpha} a_{i\gamma}sim(\alpha,\gamma),\label{eq:e4}
\end{equation}
and then rank all the unpurchased items in descending order according to their scores in Eq. (2). The top-$L$ items will be recommended to $U_i$, where $L$ is the length of recommendation list.

To compare the degree-dependent strategies, we employ a metric $H$ that counts the number of users whose recommendation lists contain the target items, say
\begin{eqnarray}
H=\sum_{i=1}^n \delta_i(R),\texttt{      } \delta_i(R)=\begin{cases} 1,&r_i\leq L \cr 0,&r_i > L \cr \end{cases},
\end{eqnarray}
where $r_i$ is the position of the target item among all $U_i$'s unpurchased items. Obviously, $0\leq H \leq (n-R)$, since the target item's degree equals $R$, and the larger value of $H$ means better performance. The number of recommended items, $L$, is limited by the user interface, with typical size no larger than 6 (see real recommendation engines of Alibaba Group and Baifendian Inc. as examples).

Since the maximum item degrees for TM and Coo8 are 617 and 933, respectively, in our simulation, we only consider $R$ ranging from 1 to 1000. To our surprising, as shown in figure 3, MaxD hardly makes new items recommended while MinD usually shows better performance. Consider the general case where the target item $O_\eta$ has established a link to user $U_i$, and $O_\alpha$ and $O_\beta$ are two of $U_i$'s collected items before $O_\eta$. For another user $U_j$ who is not connected with $O_\eta$. If $U_j$ has collected $O_\alpha$ but not $O_\beta$, then both $O_\beta$ and $O_\eta$ have the chance to be recommended to $U_j$. Since in the ICF algorithm, item similarities play the major role, let's compare the similarities $sim(\alpha,\beta)$ and $sim(\alpha,\eta)$. Statistically speaking, if $U_i$ is a very active user selected by the MaxD strategy, $O_\alpha$ and $O_\beta$ are probably less popular as indicated by the disassortative nature of the networks, therefore $k_\eta$ (i.e., $R$) may be much larger than $k_\beta$ and then $sim(\alpha,\eta)$ is probably smaller than $sim(\alpha,\beta)$, resulting in less probability of $O_\eta$ to be recommended to $U_j$. In contrast, if $U_i$ is a very inactive user selected by the MinD strategy, $O_\alpha$ and $O_\beta$ are probably of larger degrees according to the disassortative nature, resulting in smaller $sim(\alpha,\beta)$ and thus larger probability for $O_\eta$ to be recommended to $U_j$. In addition, since $U_i$ is very unpopular, it is also possible that $k_i=1$ and $U_i$ is only connected with $O_\alpha$. In such case, for all other users connected with $O_\alpha$, $O_\eta$ will be the only recommended item related to $U_i$.

\begin{figure}[!ht]
\begin{center}
\includegraphics[width=9cm]{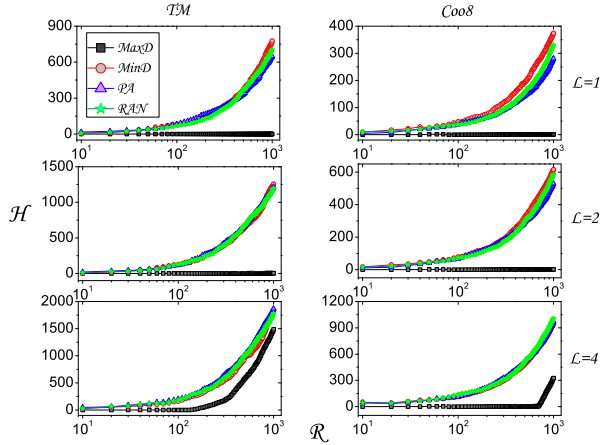}
\caption{\textbf{Performance of the four strategies for original TM and Coo8 bipartite networks.} The results of MaxD, MinD, PA and RAN are represented by black squares, red circles, blue triangles and green pentagrams, respectively. Data points are obtained by averaging over 50 independent realizations. }\label{Fig3}
\end{center}
\end{figure}

\begin{figure}[!ht]
\begin{center}
\includegraphics[width=9cm]{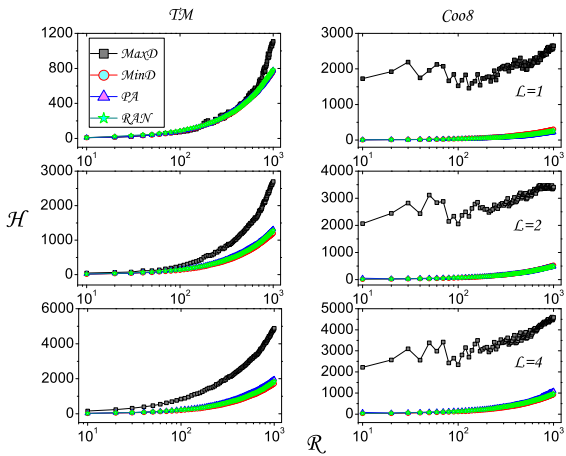}
\caption{\textbf{Performance of the four strategies for reshuffled networks.} The results of MaxD, MinD, PA and RAN are represented by black squares, red circles, blue triangles and green pentagrams, respectively. Data points are obtained by averaging over 50 independent realizations.}\label{Fig4}
\end{center}
\end{figure}

In a word, the disassortativity could contribute to the observations in figure 3. To validate this inference, we reshuffle the original networks by link-crossing method to obtain the null networks \cite{Zhangcj2012}. Specifically speaking, in each step, two links, say $(U_i,O_\alpha)$ and $(U_j,O_\beta)$, are randomly picked out, and if $U_i$ has not collected $O_\beta$ and $U_j$ has not collected $O_\alpha$, these two links are rewired as $(U_i,O_\beta)$ and $(U_j,O_\alpha)$. In one realization, we repeat such rewiring for $3w$ times. After that, the reshuffled network has identical degree sequence as the original network but the disassortative nature is vanished as shown in figure 2. Figure 4 reports the performance of the four strategies in the reshuffled networks, from which we can see that the MaxD strategy performs the best. Comparing the results for original and reshuffled networks, we conclude that the advantage of MinD strategy results from the disassortative nature of real e-commerce user-item bipartite networks. In addition, in figure 5 and figure 6, we test the performance of strategies with different $\tau$. For both TM and Coo8, the negative $\tau$ will lead to better performance while in the null networks, positive $\tau$ is better.

\begin{figure}[!ht]
\begin{center}
\includegraphics[width=9cm]{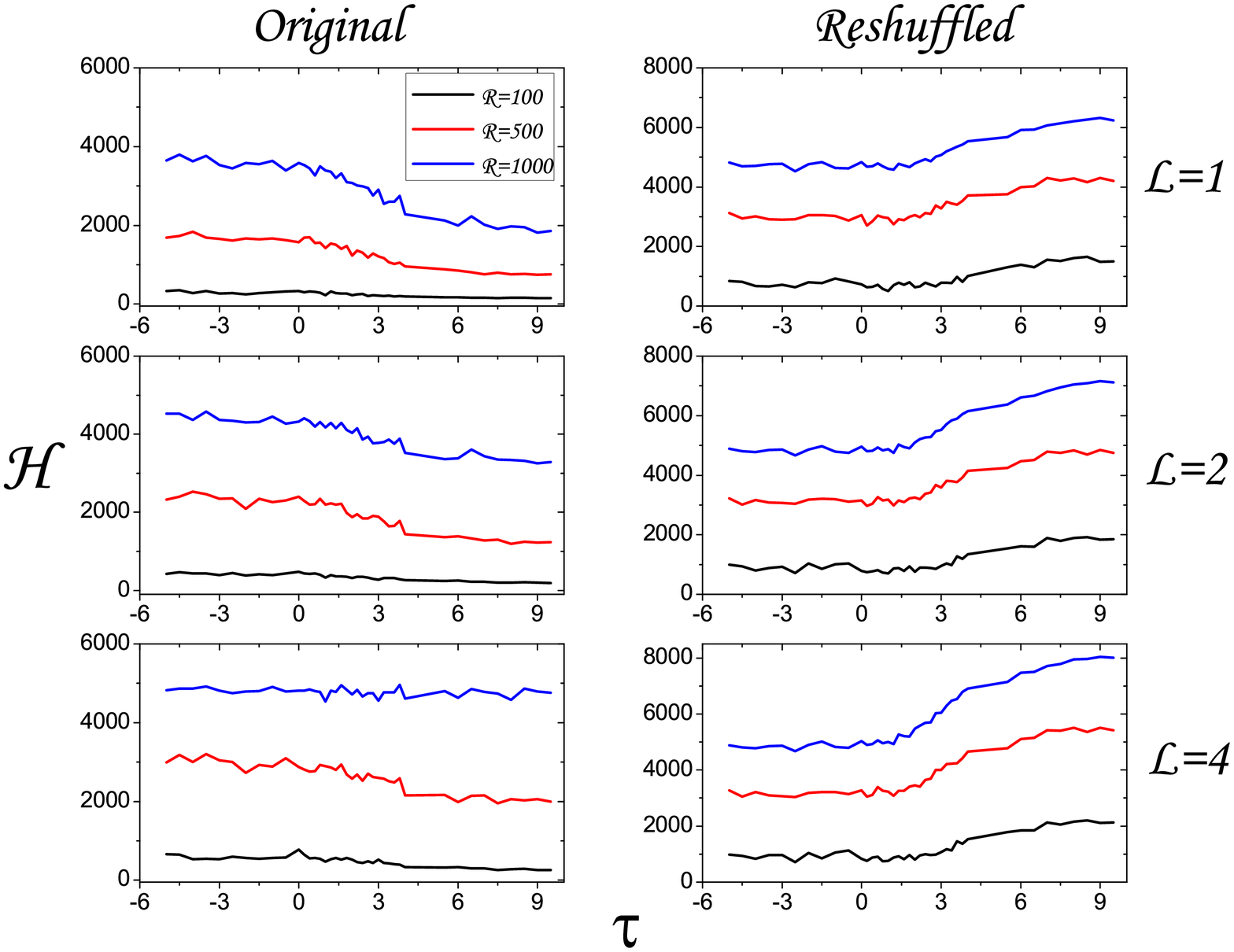}
\caption{\textbf{Performance of strategies with different $\tau$ on original and reshuffled TM networks.} The black, red and blue lines represent the results for the cases $R=100$, $R=500$ and $R=1000$, respectively. Data points are obtained by averaging over 50 independent realizations.}\label{Fig5}
\end{center}
\end{figure}

\begin{figure}[!ht]
\begin{center}
\includegraphics[width=9cm]{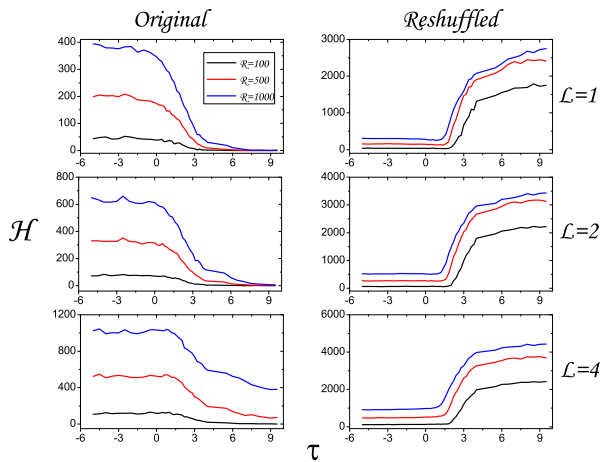}
\caption{\textbf{Performance of strategies with different $\tau$ on original and reshuffled Coo8 networks.} The black, red and blue lines represent the results for the cases $R=100$, $R=500$ and $R=1000$, respectively. Data points are obtained by averaging over 50 independent realizations.}\label{Fig6}
\end{center}
\end{figure}

\section{Discussion}
In this paper, we study a practical problem in e-commerce recommender systems: how to promote cold-start items? Under the item-based collaborative filtering systems, we show that the disassortative nature of real user-item networks leads to a non-trivial observation that to link a cold-start item to inactive users will give it more chance to appear in other users' recommendation lists. This observation is robust for varying recommendation length $L$ and linking capacity $R$. It is also applicative to some variants of item-based collaborative filtering, such as the top-$k$ nearest neighbors ICF \cite{Sarwar2001}.

Notice that, the reported results are affected by both the topological features and underlying recommendation algorithms. We have tested the user-based collaborative filtering \cite{Goldberg1992}, under which the MaxD is always better than MinD. In spite of this, this work is still relevant since in most real recommender systems, ICF plays a significant role. In addition, the perspectives and methods reported here are useful for real e-commerce applications, with the core merit is that the in-depth understanding of the structure and algorithms of recommender systems can be transferred into applicable knowledge to better market products.


\section{Acknowledgments}
We acknowledge Kuan Fang, Zhidan Zhao and Ying Zhou for useful discussions. We acknowledge Alibaba Group and Baifendian Inc. to share their real business data sets after anonymization. This work was partially supported by the National Natural Science Foundation of China under Grant Nos. 11222543. T.Z. acknowledges the Program for New Century Excellent Talents in University under Grant No. NCET-11-0070, and Special Project of Sichuan Youth Science and Technology Innovation Research Team under Grant No. 2013TD0006.


\begin{thebibliography}{99}
\bibitem{Ricci2011} F. Ricci, L. Rokach, B. Shapria, and P. B. Kantor, {\it Recommender Systems Handbook} (Springer, 2011).
\bibitem{Lv2012} L. L\"u, M. Medo, C. H. Yeung,  Y.-C. Zhang, Z.-K. Zhang, and T. Zhou, Physics Reports {\bf 519}, 1 (2012).
\bibitem{Goldberg1992} D. Goldberg, D. Nichols, B. M. Oki, and D. Terry, Communications of the ACM {\bf 35}, 61 (1992).
\bibitem{Shardanand1995} U. Shardanand, and P. Maes, In Proceedings CHI'95 Proceedings of the SIGCHI Conference on Human Factors in Computing Systems (ACM Press, New York 1995) pp. 210-217.
\bibitem{Sarwar2001} B. Sarwar, G. Karypis, J. Konstan, In Proceedings WWW'01 Proceedings of the 10th international conference on World Wide Web (ACM Press, New York 2001) pp. 285-295.
\bibitem{Deshpande2004} M. Deshpande, and G. Karypis, ACM Transactions on Information Systems {\bf 22}, 143 (2004).
\bibitem{Schafer2007} J. B. Schafer, D. Frankowski, J. Herlocker, and S. Shilad, {\it Lecture Notes in Computer Science} {Springer-Verlag, 2007}.
\bibitem{KimheungNam2010} H. N. Kim, A.-T. Ji, I. Ha, and G.-S. Jo, {\it Lecture Notes in Computer Science} {Springer-Verlag, 2010}.
\bibitem{Balabanovic1997} M. Balabanovic, and Y. Shoham, Communications of the ACM {\bf 40}, 66 (1997).
\bibitem{Pazzani2007} M. J. Pazzani, and D. Billsus, {\it Lecture Notes in Computer Science} {Springer-Verlag, 2007}.
\bibitem{Katja2013} K. Niemann, and M. Wolpers, In Proceeding KDD'13 Proceedings of the 19th ACM SIGKDD international conference on Knowledge discovery and data mining (ACM Press, New York 2013) pp. 955-963.
\bibitem{Hofmann2004} T. Hofmann, ACM Transactions on Information Systems (TOIS) {\bf 22}, 89 (2004).
\bibitem{Takacs2007} G. Tak\'acs, I. Pil\'aszy, B. N\'ameth, and D. Tikk, ACM SIGKDD Explorations Newsletter {\bf 9}, 80 (2007).
\bibitem{Bunte2012} K. Bunte, P. Schneider, B. Hammer, F.-M. Schleif, T. Villmann, and M. Biehl, Neural Networks {\bf 26}, 159 (2012).
\bibitem{Huangzan2004} Z. Huang, H. Chen, and D. Zeng, ACM Transactions on Information Systems (TOIS) {\bf 22}, 116 (2004).
\bibitem{PRE07zhoutao} T. Zhou, J. Ren, M. Medo, and Y.-C. Zhang, Physical Review E {\bf 76}, 046115 (2007).
\bibitem{PNASzhoutao} T. Zhou, Z. Kuscsik, J.-G. Liu, M. Medo, J. R. Wakeling, and Y.-C. Zhang, Proceedings of National Academy of Science of the United Sates of America {\bf 107}, 4511 (2010).
\bibitem{Zhangyc2007prl} Y.-C. Zhang, M. Medo, J. Ren, T. Zhou, T. Li, and F. Yang, Europhysics Letters {\bf 80}, 68003 (2007).
\bibitem{Zhangyc2007epl} Y.-C. Zhang, M. Blattner, and Y.-K. Yu, Physical Review Letters {\bf 99}, 154301 (2007).
\bibitem{Liujg2011} J.-G. Liu, T. Zhou, and Q. Guo, Physical Review E {\bf 84}, 037101 (2011).
\bibitem{LLY2011} L. L\"{u}, and W. Liu, Europhysics Letters {\bf 83}, 066119 (2011).
\bibitem{Zhouyanbo2013} Y. Zhou, L. L\"u, W. Liu, and J. Zhang, PLoS ONE {\bf 8}, e70094 (2013l).
\bibitem{Schein2002} A. I. Schein, A. Popescul,  L. H. Ungar, and D. M. Pennock, In Proceeding SIGIR'02 Proceedings of the 25th annual international ACM SIGIR conference on Research and development in information retrieval (ACM Press, New York 2002) pp. 253-260.
\bibitem{Par2009} S.-T Park, and W. Chu, In Proceeding Recommender Systems'09 Proceedings of the third ACM conference on Recommender systems (ACM Press, New York 2009) pp. 21-28.
\bibitem{Cantador2010} I. Cantador, A. Bellog\'{\i}n, and D. Vallet, In Proceeding Recommender Systems'10 Proceedings of the forth ACM conference on Recommender systems (ACM Press, New York 2010) pp. 237-240.
\bibitem{Ronen2013} R. Ronen, N. Koenigstein, E. Ziklik, and N. Nice, In Proceeding Recommender Systems'13 Proceedings of the 7th ACM conference on Recommender systems (ACM Press, New York 2013) pp. 407-410.
\bibitem{Zhangzk20102} Z.-K. Zhang, C. Liu, Y.-C. Zhang, and T. Zhou, Europhysics Letters {\bf 92}, 28001 (2010).
\bibitem{Zhangzk20122} Z.-K. Zhang, T. Zhou, and Y.-C. Zhang, Journal of computer science and technology {\bf 26}, 767 (2011).
\bibitem{Yin2013} D. Yin, S. Guo, B. Chidlovskii, and B. D. Davision, In Proceeding WSDM'13 Proceedings of the 6th ACM international conference on Web search and data mining (ACM Press, New York 2013) pp. 547-556.
\bibitem{Zhangl2013} L. Zhang, L.-S. Bai, and T. Zhou, Journal of University of Electronic Science and Technology of China {\bf 42}, 154 (2013).
\bibitem{Ziegler2005} C.-N. Ziegler,S. M. McNee, J. A. Konstan, and G. Lausen, In Proceeding WWW'05 Proceedings of the 14th international conference on World Wide Web (ACM Press, New York 2005) pp. 22-32.
\bibitem{Zhoutao09NJP} T. Zhou, R. Q. Su, R. R. Liu, L. L. Jiang, B. H. Wang, and Y. C Zhang, New Journal of Physics {\bf 11}, 123008 (2011).
\bibitem{Zhoujunlin2011} J.-L Zhou, Y. Fu, H. Lu, and C.-J Sun, Journal of Computer Science and Technology {\bf 26}, 816 (2011).
\bibitem{Lam2004} S. K. Lam, and J. Riedl, In Proceeding WWW'04 Proceedings of the 13th international conference on World Wide Web (ACM Press, New York 2004) pp. 393-402.
\bibitem{Mobasher2005} B. Mobasher, R. Burke, R. Bhaumik, and C. Williams, In Proceedings of the 2005 WebKDD Workshop (ACM Press, New York 2005).
\bibitem{Mobasher2006} B. Mobasher, R. Burke, R. Bhaumik, and C. Williams, {\it Lecture Notes in Computer Science} {Springer-Verlag, 2006}.
\bibitem{Mehta2007} B. Mehta, T. Hofmann, and P. Fankhauser, In Proceeding IUI'07 Proceedings of the 12th international conference on Intelligent user interfaces (ACM Press, New York 2007) pp. 14-21.
\bibitem{Mobasher2007} B. Mobasher, R. Burke, R. Bhaumik, and C. Williams, ACM Transactions on Internet Technology {\bf 7}, 1 (2007).
\bibitem{Verhoef2009} P. C. Verhoef, and P. S. H. Leeflang, Journal of Marketing {\bf 73}, 14 (2009).
\bibitem{Edward2008} F. C. Edward, C. J. Hadlock, and J. R. Pierce, The Review of Financial Studies {\bf 22}, 2361 (2008).
\bibitem{lvlinyuan2011} L. L\"u, and T. Zhou, Physica A {\bf 390}, 1150 (2011).
\bibitem{Huang2007} Z. Huang, D. D. Zeng, and H. Chen, Management Science {\bf 53}, 1146 (2007).
\bibitem{Koren2011} Y. Koren, and J. Sill, In Proceedings of the 5th ACM Conference on Recommender Systems (ACM Press, New York 2011) pp. 117-124.
\bibitem{Zhangqm2013} Q.-M. Zhang, A. Zeng, and M.-S. Shang, PLoS ONE {\bf 8}, e62624 (2013).
\bibitem{Weizeng2014} W. Zeng, A. Zeng, H. Liu, M.-S. Shang, and T. Zhou, arXiv 1402.6132 (2013).
\bibitem{Barabasi199901} A.-L. Barab\'{a}si, and R. Albert, Science {\bf 286}, 509 (1999).
\bibitem{Lambiotte2005} R. Lambiotte, and M. Ausloos, Physical Review E {\bf 72}, 066107 (2005).
\bibitem{Goncalves2008} B. Goncalves, and J. J. Ramasco, Physical Review E {\bf 78}, 026123 (2008).
\bibitem{Zimo2012} Z. Yang, Z.-K. Zhang, and T. Zhou, Europhysics Letters {\bf 100}, 68002 (2012).
\bibitem{Shangmz2010} M.-S. Shang, L. L\"u, Y.-C. Zhang, and T. Zhou, Europhysics Letters {\bf 90}, 48006 (2010).
\bibitem{Poster2001} R. Pastor-Satorras, A. V\'{a}zquez, and A. Vespignani, Physical Review Letters {\bf 87}, 258701 (2001).
\bibitem{Zhangcj2012} C.-J. Zhang, and A. Zeng, Physica A {\bf 391}, 1822 (2012).
\bibitem{Goldstein2004} M. L. Goldstein, S. A. Morris, and G. G. Yen, European Physical Journal B {\bf 41}, 255 (2004).
\bibitem{Clauset2009} A. Clauset, C. R. Shalizi, and M. E. J. Newman, SIAM Review {\bf 51}, 661 (2009).

\end{thebibliography}
\end{document}